\documentclass[pra,aps,groupaddress,showpacs]{revtex4}
\usepackage{epsfig,amsmath,graphicx,amssymb,overpic}
\usepackage{bm}
\usepackage{mathrsfs}
\usepackage{graphicx}
\usepackage{epsfig}
\usepackage{dcolumn}
\usepackage{bm}
\usepackage{amsmath,amssymb,amsthm}
\usepackage[colorlinks=true,linkcolor=blue]{hyperref}
\usepackage{subfigure}
\usepackage{booktabs}
\usepackage[mathscr]{euscript}
\usepackage{ulem}
\usepackage{caption}
\usepackage{ragged2e}

\newcommand{\bea}{\begin{eqnarray}}
\newcommand{\eea}{\end{eqnarray}}
\newcommand{\beq}{\begin{equation}}
\newcommand{\eeq}{\end{equation}}

\def\/{\over}

\begin{document}

\title{Deformation Conjecture: Deforming Lower Dimensional Integrable Systems to Higher Dimensional Ones by Using Conservation Laws}
\author{S. Y. Lou$^1$, Xia-zhi Hao$^2$ and Man Jia$^1$}
\affiliation{ \footnotesize{$^1$School of Physical Science and Technology, Ningbo University, Ningbo, 315211, China}\\
\footnotesize{$^2$Faculty of Science, Zhejiang University of Technology, Hangzhou, 310014, China}}
\begin{abstract}
Utilizing some conservation laws of $(1+1)$-dimensional integrable local evolution systems,  it is conjectured that higher dimensional integrable equations may be regularly constructed by a deformation algorithm. The algorithm can be applied to Lax pairs and higher order flows. In other words, if the original lower dimensional model is Lax integrable (possesses Lax pairs) and symmetry integrable (possesses infinitely many higher order symmetries and/or infinitely many conservation laws), then the deformed higher order systems are also Lax integrable and symmetry integrable. For concreteness, the deformation algorithm is applied to the usual $(1+1)$-dimensional Korteweg-de Vries (KdV) equation and the $(1+1)$-dimensional Ablowitz-Kaup-Newell-Segur (AKNS) system (including nonlinear Schr\"odinger (NLS) equation as a special example). It is interesting that the deformed (3+1)-dimensional KdV equation is also an extension of the $(1+1)$-dimensional Harry-Dym (HD) type equations which are reciprocal links of the (1+1)-dimensional KdV equation. The Lax pairs of the $(3+1)$-dimensional KdV-HD system and the $(2+1)$-dimensional AKNS system are explicitly given. The higher order symmetries, i.e., the whole $(3+1)$-dimensional KdV-HD hierarchy, are also explicitly obtained via the deformation algorithm. The single soliton solution of the $(3+1)$-dimensional KdV-HD equation is implicitly given. Because of the effects of the deformation, the symmetric soliton shape of the usual KdV equation is no longer conserved and deformed to be asymmetric and/or multi-valued. The deformation conjecture holds for all the known $(1+1)$-dimensional integrable local evolution systems that have been checked, and we have not yet found any counter-example so far. The introduction of a large number of ($D+1$)-dimensional integrable systems of this paper explores a serious challenge to all mathematicians and theoretical physicists because the traditional methods are no longer directly valid to solve these integrable equations.
\end{abstract}

\maketitle

\bf Introduction. \rm
Integrable systems and/or soliton equations are widely applied in almost all of the modern physical fields including the condensed matter physics \cite{CMP,Sci}, particle and nuclear physics \cite{PNP}, field theory \cite{FT}, universe \cite{Univ}, fluid mechanics \cite{FM}, geophysics \cite{GeoP}, planetary and space science \cite{PSS}, optics \cite{Opt}, plasma physics \cite{Plsm}, and in other scientific and technological fields such as biology \cite{Bio}, communications \cite{Commun}, chemistry \cite{Chem}, etc.
Usually, the typical soliton equations are in (1+1)-dimensional cases. To extend (1+1)-dimensional soliton equations to higher dimensions is one of the important and significant topics both in mathematics and physics \cite{Fofas,hevenly,dual,Ptest}. In this manuscript, we try to find higher dimensional integrable systems from lower dimensional ones by proposing a deformation algorithm.

Contracting a complex theory to a simpler one is not a difficult work. For instance, the limiting procedures $\hbar\rightarrow 0$ and $c\rightarrow 0$ (where $\hbar$ is the Planck constant and $c$ is the speed of light) will render a relativistic quantum theory degenerating into a classical theory. However, the inverse procedure, i.e., the deformation of a simple theory to a more complex theory is very difficult. Fortunately, there are some unusual methods to deal with the inverse problem. In some special circumstances, it is possible to accomplish the inverse procedure. For example, the classical Yang-Baxter equation may evolve to the quantum Yang-Baxter equation. Some critical phenomenon theory treated by conformal field theory, where the mass of the system is zero, may be deformed to a theory for the nonzero mass case. Would such a problem occur also in nonlinear equations? That is to say could a solution of a complex equation be obtained from a simple equation? In deed, some special solutions of the well known sine-Gordon (SG) equation in $D+1$ dimensions can be deformed to those of the doubled SG (DSG) equation \cite{DsG,DsG1}. Solutions of the $\phi^4$ model may also be deformed to those of the $\phi^6+\phi^4$ model and the $\phi^3+\phi^4$ model (or named T. D. Lee model) \cite{phi4,phi41}.

To reduce a high dimensional integrable system to lower dimensional ones, one can use some different approaches like the classical and nonclassical Lie approaches \cite{Symm,Symm1,Symm2}, symmetry constraints \cite{SymC,SymD} and decompositions \cite{Dec,Cao}. From the studies of the (1+1)- and (2+1)-dimensional integrable models, one knows that there are several ordinary differential equation (ODE) reductions such as the Riccati equation and the Painlev\'e I-VI equations \cite{Pain} for all known integrable lower dimensional partial differential equations (PDEs). This fact shows that all of the known integrable models may be considered as different deformations of some ODEs. For instance, some (1+1)- and (2+1)-dimensional integrable sine-Gordon and Tzitzeca equations can be considered as the deformations of a simple Riccati equation via Miura type deformation relations \cite{Riccati}. Here, we propose a universal algorithm to deform lower dimensional integrable systems to higher dimensional ones via a deformation conjecture. For the examples given below, a central role is played by the following conjecture.

\bf Deformation conjecture. \it For a general (1+1)-dimensional integrable local evolution system
\begin{equation}
u_t=F(u,\ u_x,\ \ldots,\ u_{xn}),\ u_{xn}=\partial_x^nu, u=(u_1,\ u_2,\ \ldots,\ u_m), \label{Gen}
\end{equation}
if there exist some conservation laws
\begin{equation}
\rho_{it}=J_{ix}, \ i=1,\ 2,\ \ldots, D-1,\ \rho_i=\rho_i(u),\ J_i=J_i(u,\ u_x,\ \ldots,\ u_{xN}), \label{CLaw}
\end{equation}
where the conserved densities $\rho_i$ are dependent only on the field $u$ while the flows $J_i$ can be field derivative dependent, then the deformed $(D+1)$-dimensional system
\begin{equation}
\hat{T}u=F(u,\ \hat{L}u,\ \ldots,\ \hat{L}^nu)\ \label{D+1}
\end{equation}
may be integrable with the deformation operators
\begin{equation}
\hat{L}\equiv \partial_x+\sum_{i=1}^{D-1}\rho_i\partial_{x_i},\
\hat{T}\equiv \partial_t+\sum_{i=1}^{D-1}\bar{J}_i\partial_{x_i} \label{LT}
\end{equation}
and the deformed flows
\begin{equation}
\bar{J_i}=\left.J_i\right|_{u_{xj}\rightarrow \hat{L}^ju,\ j=1,\ 2,\ \ldots,\ N}. \label{Jbar}
\end{equation}
\rm

In accordance with the deformation conjecture, any $(1+1)$-dimensional integrable system possessing conservation laws will result in a higher dimensional integrable one. Though the conjecture has not yet been strictly proven, we have already checked the correctness, effectiveness and flexibility of the conjecture for almost all the known $(1+1)$-dimensional local integrable evolution systems such as the Korteweg-de Vries (KdV) equation \cite{KdV}, the modified KdV equation \cite{mKdV}, the Harry-Dym equation \cite{HD}, the Burgers equation \cite{BE}, the Sharma-Tasso-Olver equation \cite{STO}, the Ablowitz-Kaup-Newell-Segur (AKNS) system (including the nonlinear Schr\"odinger equation (NLS)) \cite{AKNS,AKNS1}, the fifth order KdV equation \cite{5KdV}, the Sawada-Kotera equation \cite{SK}, the Kaup-Kupershmidt equation \cite{KK}, the Boussinesq equation \cite{Bq}, the dispersive long wave system \cite{DLWE}, the Ito system \cite{Ito}, the Hirota-Satsuma system \cite{HS}, the Kaup-Broer system \cite{KB},  the Kaup-Newell system (including the first type of derivative NLS equation) \cite{KN}, the Chen-Lee-Liu system (including the second type of derivative NLS equation) \cite{CLL,CLL1}, the Gerdjikov-Ivanov equation (including the third type of derivative NLS equation) \cite{GI,GI1}, the Landau-Lifshitz equation (including the Heisenberg equation) \cite{LL,LL1}, the Levi system \cite{Levi}, the Manakov system \cite{Mana}, and so on.

We should mention that discrete integrable systems, such as Toda lattice systems, Calogero-Moser-Sutherland systems, etc., have not been considered here. The deformation conjecture for these discrete integrable systems should be excluded and need further consideration.

\bf Higher dimensional KdV-HD equations. \rm We first examine the algorithm for constructing higher dimensional integrable systems on a classical example \cite{KdV}, the KdV equation
\begin{equation}
u_t=u_{xxx}+6uu_x. \label{KdV}
\end{equation}
The KdV equation \eqref{KdV} was first introduced by Boussinesq \cite{Bq} and rediscovered by Diederik Korteweg and Gustav de
Vries \cite{KdV}. It has various connections to physical problems, which approximately describes the evolution of long, one-dimensional waves in many physical settings, including shallow-water waves with weak nonlinear
restoring forces, long internal waves in a density-stratified ocean, ion acoustic waves in a plasma, acoustic waves
on a crystal lattice \cite{KdVappl} and the two-dimensional quantum gravity \cite{gravity}. The KdV equation can be solved using the inverse scattering transform and other methods such as those applied to other integrable systems \cite{IST}.

There are infinite number of conservation laws satisfied by solutions to the KdV equation \eqref{KdV}. Only two will be of concern to us
\begin{eqnarray}
&&\rho_{1t}=J_{1x},\ \rho_1=u,\ J_1=u_{xx}+3u^2,\label{CLKdV1}\\
&&\rho_{2t}=J_{2x},\ \rho_2=u^2,\ J_2=2uu_{xx}-u_x^2+4u^3.\label{CLKdV2}
\end{eqnarray}
A simple comparison with deformation conjecture leads to the discovery of a (3+1)-dimensional KdV-HD equation
\begin{eqnarray*}
\hat{T}u=\hat{L}(\hat{L}^2u+3u^2),
\end{eqnarray*}
or equivalently,
\begin{eqnarray}
u_t=\hat{L}(\hat{L}^2u+3u^2)-u_y\bar{J}_1-u_z\bar{J}_2\equiv \bar{K}_3,\label{3+1KdV}
\end{eqnarray}
where the deformed operators $\hat{L}$ and $\hat{T}$ possess the forms
\begin{eqnarray}
\hat{L}=\partial_x+u\partial_y+u^2\partial_z,\ \hat{T}=\partial_t+\bar{J}_1\partial_y+\bar{J}_2\partial_z\label{KdVLT}
\end{eqnarray}
with the deformed flows $\bar{J}_1=\hat{L}^2u+3u^2$ and $\bar{J}_2=2u\hat{L}^2u-(\hat{L}u)^2+4u^3$.

To emphasize the non-triviality of the (3+1)-dimensional KdV-HD equation \eqref{3+1KdV}, we impose some reductions to this equation and derive from it some of known and novel integrable equations in (1+1)- and (2+1)-dimensions.
Equation \eqref{3+1KdV} reduces to the usual (1+1)-dimensional KdV equation \eqref{KdV} provided that $u$ is independent of $y$ and $z$.

If $u$ is $\{x,\ z\}$-independent, then \eqref{3+1KdV} becomes
\begin{eqnarray}
u_t=\left(u^3u_{yy}+u^3\right)_y, \label{HD}
\end{eqnarray}
which is equivalent to another well known (1+1)-dimensional physical model, the Harry-Dym (HD) equation \cite{HD1}. In other words, \eqref{HD} is a reciprocal link of the KdV equation \eqref{KdV} related to the conservation law \eqref{CLKdV1}. It is not difficult to find that the Harry-Dym equation possesses two conservation laws $\rho^{\pm}_t=J^{\pm}_y$ with $\rho^{\pm}=u^{\pm 1},\ J^+=u^3(u_{yy}+1)$ and $J^-=-(uu_y)_y-3u$. Applying the deformation conjecture to \eqref{HD} with these two conservation laws, one can re-obtain the (3+1)-dimensional KdV-HD equation \eqref{3+1KdV}.

If $u$ is $\{x,\ y\}$-independent, then \eqref{3+1KdV} becomes a new reciprocal link of the KdV equation \eqref{KdV}
\begin{eqnarray}
u_t=\left(u^6u_{zz}+\frac12u^4\right)_z+3u^4u_z^3\label{HD2}
\end{eqnarray}
related to the conservation law \eqref{CLKdV2}. The model \eqref{HD2} possesses the conservation laws $(u^{-1})_t=(-u^4u_{zz}-u^3u_z^2-u^2)_z$ and $(u^{-2})_t=(-2u^3u_{zz}-3u^2u_z^2-4u)_z$. The application of the deformation conjecture and these two conservation laws to the model \eqref{HD2} will again lead to the (3+1)-dimensional KdV-HD equation \eqref{3+1KdV}.

When $u$ is $z$-independent, \eqref{3+1KdV} becomes a (2+1)-dimensional KdV-HD equation
\begin{eqnarray}
u_t=\left(u_{xx}+3u^2+3uu_{xy}+\frac32u^2u_{yy}\right)_x
+\left(u^3u_{yy}+u^3+\frac32u^2u_{xy}\right)_y,\label{2+1KdV}
\end{eqnarray}
which is an integrable (2+1)-dimensional combination of the KdV equation \eqref{KdV} and its reciprocal link \eqref{HD}.

When $u$ is $y$-independent, \eqref{3+1KdV} becomes another (2+1)-dimensional KdV-HD equation
\begin{eqnarray}
u_t=\left(u_{xx}+3u^2+3u^2u_{xz}-6u^3u_{z}^2\right)_x
+\left(u^6u_{zz}+\frac12u^4+3u^4u_{xz}+6u^3u_xu_z\right)_z+3u_z(u_x+u^2u_z)^2,\label{2+1KdV1}
\end{eqnarray}
which is an integrable (2+1)-dimensional combination of the KdV equation \eqref{KdV} and its reciprocal link \eqref{HD2}.

When $u$ is $x$-independent, \eqref{3+1KdV} becomes an integrable (2+1)-dimensional combination of two different reciprocal links \eqref{HD} and \eqref{HD2}
\begin{eqnarray}
u_t=\left[u^3(1+u_{yy}+3u^2u_{zz})+9u^3u_zv_-\right]_y
+\left[\frac12u^4(2u^2u_{zz}+6u_{yy}+1)-9u^3u_yv_-\right]_z+3u^2u_zv_+^2,\label{2+1KdV2}
\end{eqnarray}
where $v_{\pm}=uu_z\pm u_y$.

When one higher dimensional equation is obtained from a lower dimensional one via deformation conjecture, the corresponding integrable properties may also be transferred from the original equation by the similar deformation algorithm. To prove the Lax integrability of the (3+1)-dimensional KdV equation \eqref{3+1KdV}, one can apply the similar deformation algorithm to the Lax pair of the original (1+1)-dimensional KdV equation \eqref{KdV}
\begin{eqnarray}
&& (\partial_{x}^2+u-\lambda)\psi\equiv {M}\psi=0,\nonumber\\
&& (\partial_t-4\partial_x^3-6u\partial_x-3u_x)\psi\equiv {N}\psi=0.\label{Lax}
\end{eqnarray}
The compatibility condition $[{M},\ {N}]={M} {N}-{N}{M}=0$ is equivalent to the KdV equation \eqref{KdV}.

It is interesting that applying the deformation relations
\begin{equation}
\partial_x\rightarrow \hat{L},\ \partial_t\rightarrow \hat{T} \label{deform}
\end{equation}
on \eqref{Lax}, the Lax pair of the (3+1)-dimensional KdV equation \eqref{3+1KdV} can then be easily put in the form
\begin{eqnarray}
&& (\hat{L}^2+u-\lambda)\psi\equiv \hat{M}\psi=0,\nonumber\\
&& \big[\hat{T}-4\hat{L}^3-6u\hat{L}-3(\hat{L}u)\big]\psi\equiv \hat{N}\psi=0.\label{Lx}
\end{eqnarray}

Now, it is straightforward to verify that the linear problem \eqref{Lx} is just the Lax pair of the (3+1)-dimensional KdV equation \eqref{3+1KdV}. In other words, the compatibility condition of \eqref{Lx}
\begin{eqnarray}
[\hat{M},\ \hat{N}]=\hat{M}\hat{N}-\hat{N}\hat{M}=0\label{MN}
\end{eqnarray}
gives the (3+1)-dimensional KdV equation \eqref{3+1KdV}.

It is also known that the (1+1)-dimensional KdV equation is symmetry integrable that means \eqref{KdV} possesses infinitely many higher order symmetries $K_{2n+1}=\Phi^n u_x$. In other words, there are infinitely many higher order commuting flows
\begin{equation}
u_{t_{2n+1}}=K_{2n+1}=\Phi^nu_x,\ \Phi=\partial_x^2+4u+2u_x\partial_x^{-1}, \ n=0,\ 1,\ \ldots,\ \infty, \label{KdVH}
\end{equation}
which is called as the KdV hierarchy. For $n=1$, \eqref{KdVH} is just the KdV equation ($t_3=t$) while the fifth order KdV equation
\begin{equation}
u_{t_{5}}=(u_{x4}+10uu_{xx}+5u_x^2+10u^3)_x \label{KdV5}
\end{equation}
is related to \eqref{KdVH} with $n=2$.
It is known that $u$ and $u^2$ are also conserved densities for the whole KdV hierarchy \eqref{KdVH} with the differential polynomial flows in the forms
\begin{eqnarray}
u_{t_{2n+1}}&=&(J_{2n+1})_x,\ J_{2n+1}=\partial_x^{-1}\Phi^nu_x,\ \nonumber\\
(u^2)_{t_{2n+1}}&=&(G_{2n+1})_x,\ G_{2n+1}=2\partial_x^{-1}u\Phi^nu_x,\  \label{JGn}
\end{eqnarray}
where $J_{2n+1}$ and $G_{2n+1}$ are differential polynomials of $u$ with respect to $x$. For $n=2$,
\begin{eqnarray}
 J_{5}&=&u_{x4}+10uu_{xx}+5u_x^2+10u^3, \nonumber\\
G_{5}&=&2uu_{x4}-2u_xu_{x3}+u_{xx}^2+20u^2u_{xx}+15u^4.\label{JG5}
\end{eqnarray}
Owing to the conservation laws \eqref{JGn} and the deformation assumption, we can obtain the whole (3+1)-dimensional KdV hierarchy
\begin{eqnarray}
\hat{T}_{2n+1}u=\left.\left(\Phi^nu_x\right)\right|_{u_{xm}\rightarrow \hat{L}^mu,\ m=1,\ 2,\ \ldots,\ 2n+1},\  \label{KdVH3}
\end{eqnarray}
where $ \hat{T}_{2n+1}=\partial_{t_{2n+1}}+\bar{J}_{2n+1}\partial_y+\bar{G}_{2n+1}\partial_z, \hat{L}=\partial_x+u\partial_y+u^2\partial_z,\ \bar{J}_{2n+1}=\left.J_{2n+1}\right|_{u_{xm}\rightarrow \hat{L}^mu,\ m=1,\ 2,\ \ldots,\ 2n+1} $ and
$ \bar{G}_{2n+1}=\left.G_{2n+1}\right|_{u_{xm}\rightarrow \hat{L}^mu,\ m=1,\ 2,\ \ldots,\ 2n+1}$.
The (3+1)-dimensional KdV hierarchy \eqref{KdVH3} is equivalent to
\begin{equation}
u_{t_{2n+1}}=\left.\left(1-u_y\partial_x^{-1}-2u_z\partial_x^{-1}u\right)
\Phi^nu_x\right|_{u_{xm}\rightarrow
\hat{L}^mu,\ m=1,\ 2,\ \ldots,\ 2n+1} \equiv \bar{K}_{2n+1} \label{3KdVH}
\end{equation}
with the first three $\bar{K}_{2n+1}$ being $\bar{K}_1=u_x$, $\bar{K}_3=(\hat{L}-u_y)(\hat{L}^2u+3u^2)-u_z\left[2u\hat{L}^2u-(\hat{L}u)^2+4u^3\right]$ and
$$\bar{K}_5=(\hat{L}-u_y)\left[\hat{L}^4u+10u\hat{L}^2u+5(\hat{L}u)^2+10u^3\right]
-u_z\left[2u\hat{L}^4u-2(\hat{L}u)(\hat{L}^3u)+(\hat{L}^2u)^2+20u^2\hat{L}^2u+15u^4\right].$$
One can check that the compatibility conditions, $u_{t_{2n+1}t_{2m+1}}=u_{t_{2m+1}t_{2n+1}}$, of the (3+1)-dimensional KdV hierarchy \eqref{3KdVH} are satisfied, i.e.,
\begin{equation}
[\bar{K}_{2n+1},\ \bar{K}_{2m+1}]=\lim_{\epsilon\rightarrow 0} \frac{\mbox{d}}{\mbox{d}\epsilon}\left[\bar{K}_{2n+1}(u+\epsilon \bar{K}_{2m+1})- \bar{K}_{2m+1}(u+\epsilon \bar{K}_{2n+1})\right]=0.
\end{equation}

\bf The (2+1)-dimensional AKNS system. \rm In additional to the KdV equation, the NLS equation is another important physical model to describe various physical phenomena including Bose-Einstein condensates \cite{BEC}, quantum spin \cite{Spin}, quantum optics \cite{Opt1},
nonequilibrium physics in biology \cite{Bio1}, string theory and gravity \cite{string}, particle physics \cite{particle}, oceanic waves \cite{Ocean},
nonlinear optics \cite{NLOpt}, plasma physics \cite{plasma}, atmospheric
science \cite{atmos}, superfluids \cite{sfluid}, acoustic waves \cite{acous}, economics \cite{yan}, and so on. The more general form of the NLS equation is the so-called AKNS system
\begin{equation}
u_t=u_{xx}+2u^2v,\ v_t=-v_{xx}-2v^2u. \label{AKNS}
\end{equation}
The usual NLS equation is related to the AKNS system \eqref{AKNS} by $v=u^*$ and $t\rightarrow \sqrt{-1}t$. Generally, the AKNS system is believed to contain certain important classes of evolution equations \cite{AKNS1} which can be solved by the inverse scattering method.

We illustrate the algorithm for the case of the AKNS system.
According to the deformation conjecture, to deform the AKNS system \eqref{AKNS} to a higher dimensional integrable one, we have to find some conservation laws of \eqref{AKNS}. Fortunately, it is known that the AKNS system \eqref{AKNS} possesses the conservation law
\begin{equation}
(uv)_t=(vu_x-uv_x)_x.  \label{CLAKNS}
\end{equation}
Thus, applying the deformation conjecture, we can obtain a deformed (2+1)-dimensional AKNS system
\begin{eqnarray}
&&\hat{T}u=\hat{L}^2u+2u^2v,\ \hat{T}v=-\hat{L}^2v-2v^2u,  \label{2AKNS}\\
&&\hat{L}=\partial_x+uv\partial_y,\ \hat{T}=\partial_t+\left[v(\hat{L}u)-u(\hat{L}v)\right]\partial_y, \nonumber
\end{eqnarray}
i.e.,
\begin{eqnarray}
&&u_t=u_{xx}+2u^2v+u^2v(vu_{yy}+2u_yv_y)+2u(vu_{xy}+v_xu_y),\nonumber\\
&&v_t=-v_{xx}-2v^2u-uv^2(uv_{yy}+2u_yv_y)-2v(uv_{xy}+u_xv_y). \label{2AKNSuv}
\end{eqnarray}
Similar to the (3+1)-dimensional KdV equation, the (2+1)-dimensional AKNS system \eqref{2AKNSuv} is also Lax integrable and symmetry integrable.
We can go directly to the Lax pair of the (2+1)-dimensional AKNS system \eqref{2AKNSuv}
\begin{equation}
\left(\begin{array}{c} \psi_1 \\ \psi_2\end{array}\right)_x=-uv\left(\begin{array}{c} \psi_1 \\ \psi_2\end{array}\right)_y+\left(\begin{array}{cc} \lambda & -v \\ u & -\lambda \end{array}\right)\left(\begin{array}{c} \psi_1 \\ \psi_2\end{array}\right), \label{LaxAKNS}
\end{equation}
\begin{equation}
\left(\begin{array}{c} \psi_1 \\ \psi_2\end{array}\right)_t=(uv_x-vu_x+u^2vv_y-v^2uu_y)\left(\begin{array}{c} \psi_1 \\ \psi_2\end{array}\right)_y+\left(\begin{array}{cc} -2\lambda^2-uv &v_x+uvv_y+2\lambda v \\ u_x+vuu_y-2\lambda u & 2\lambda^2+uv \end{array}\right)\left(\begin{array}{c} \psi_1 \\ \psi_2\end{array}\right) \label{LatAKNS}
\end{equation}
by repeating the algorithm that we followed previous for the (3+1)-dimensional KdV equation.

When the fields $\psi_1,\ \psi_2,\ u$ and $v$ are $y$-independent, the Lax pair \eqref{LaxAKNS}-\eqref{LatAKNS} of the (2+1)-dimensional AKNS system \eqref{2AKNSuv} is reduced back to the Lax pair of the usual (1+1)-dimensional AKNS system \eqref{AKNS}.

When the fields $u$ and $v$ of \eqref{2AKNSuv} are $y$ independent, the (2+1)-dimensional AKNS system is return back to the usual AKNS equations. For $u_x=v_x=0$, the (2+1)-dimensional AKNS system \eqref{2AKNSuv} becomes a reciprocal link related to the conservation law \eqref{CLAKNS}
\begin{eqnarray}
&&u_t=u^2v(vu_{yy}+2u_yv_y)+2u^2v,\nonumber\\
&&v_t=-uv^2(uv_{yy}+2u_yv_y)-2v^2u. \label{AKNSy}
\end{eqnarray}
The reciprocal link \eqref{AKNSy} of the AKNS system possesses the following conservation law
\begin{eqnarray}
(u^{-1}v^{-1})_t=(uv_y-vu_y)_y. \label{CLAKNSy}
\end{eqnarray}

By using the deformation conjecture and the conservation law \eqref{CLAKNSy} to the model \eqref{AKNSy}, one can re-obtain the $(2+1)$-dimensional AKNS system \eqref{2AKNSuv}.

\bf Summary and discussions. \rm In summary, applying the deformation conjecture to the known (1+1)-dimensional integrable evolution models with $D-1$ conservation laws, $[\rho_j(u)]_t=[J_j(u,\ u_x,\ \ldots)]_x$, one can obtain many $(D+1)$-dimensional integrable systems. The fact that the deformation operation that distills the higher dimensional equations from the lower dimensional ones which possesses the Lax and symmetry integrability, in some sense,  preserves integrability, implies these higher dimensional equations be Lax integrable and symmetry integrable. The deformation algorithm will destroy the Painlev\'e integrability because the reciprocal transformations are included. The Lax pairs of the $(3+1)$-dimensional KdV-HD equation \eqref{3+1KdV} and the $(2+1)$-dimensional AKNS \eqref{2AKNSuv} are explicitly given.
The deformation conjecture has the advantage that it constructs $(D+1)$-dimensional integrable systems, as well as integrable hierarchies, in a straightforward manner.
The whole $(3+1)$-dimensional KdV  hierarchy has also been obtained via the deformation algorithm from the $(1+1)$-dimensional KdV hierarchy.

Usually, ($n+1$)-dimensional integrable models for $n\geq 2$ such as the Kadomtsev-Petviashivili equation and the Davey-Stewartson equation are nonlocal in their evolution forms. However, the ($D+1$)-dimensional integrable models obtained from the deformation algorithm are all local.

For every conservation law, $\rho_t=J_x$, of (1+1)-dimensional local evolution systems, the deformation algorithm included in the deformation conjecture implies a reciprocal transformation of the original $(1+1)$-dimensional model by deformation relations $\partial_x\rightarrow \rho\partial_y$ and $\partial_t\rightarrow \partial_t+\bar{J}\partial_y$.

Though the deformation conjecture requires the conserved density is only the field $u$-dependent but not the $x$-derivatives of the field, one can make suitable transformations such that the field derivative dependent conserved density becomes the field derivative independent for the new models. For instance, the AKNS system \eqref{AKNS} possesses a conservation law with the conserved density $vu_x$. By using the dependent variable transformations, $vu_x=u_1,\ uv=u_2$, the AKNS system \eqref{AKNS} is transformed to a variant form of the AKNS system (VAKNS)
\begin{equation}
u_{1t}=[u_2^2+u_{1x}+2u_1u_2^{-1}(u_1-u_{2x})]_x,\ u_{2t}=(2u_1-u_{2x})_x. \label{VAKNS}
\end{equation}
 The VAKNS system \eqref{VAKNS} possesses three field derivative independent conserved densities, $u_1,\ u_2$ and $u_1u_2^{-1}$ with $(u_1u_2^{-1})_t=(u_2^{-1}u_{1x}-u_1u_2^{-2}u_{2x}+u_1^2u_2^{-2}+2u_2)_x$. Thus, the VAKNS system can be deformed to a (4+1)-dimensional integrable local evolution system by means of the deformation algorithm,
\begin{equation}
\hat{T}u_{1}=\hat{L}[u_2^2+\hat{L}u_{1}+2u_1u_2^{-1}(u_1-\hat{L}u_{2})],\
\hat{T}u_{2}=\hat{L}(2u_1-\hat{L}u_{2})\label{4VAKNS}
\end{equation}
with $\hat{L}=\partial_x+u_1\partial_y+u_2\partial_z+u_1u_2^{-1}\partial_{\xi}$, $\hat{T}=\partial_t+\bar{J}_1\partial_y+\bar{J}_2\partial_z+\bar{J}_3\partial_{\xi}$, and $\{\bar{J}_1=u_2^2+\hat{L}u_{1}+2u_1u_2^{-1}(u_1-\hat{L}u_{2}),\ \bar{J}_2=2u_1-\hat{L}u_{2},\ \bar{J}_3=u_2^{-1}\hat{L}u_{1}-u_1u_2^{-2}\hat{L}u_{2}+u_1^2u_2^{-2}+2u_2\}$. The (4+1)-dimensional VAKNS system \eqref{4VAKNS} is a combination of the VAKNS system \eqref{VAKNS} and its three reciprocal links related to the conserved densities, $u_1$, $u_2$ and $u_1u_2^{-1}$, respectively.

In fact, for the AKNS system \eqref{AKNS}, there are some other important variant forms such as the KN system \cite{KN}, the Levi system \cite{Levi}, the Heisenberg system \cite{LL1}, and so on. For these variant forms, there are three or four field derivative independent conserved densities. Therefore, some (4+1)- and (5+1)-dimensional integrable local systems can be found by applying the deformation conjecture on these variant forms.

It should be mentioned that for any ($D+1$)-dimensional deformed integrable systems, there are some new (1+1)-dimensional reductions. Applying the deformation algorithm on these reductions, one can re-discover the ($D+1$)-dimensional deformed integrable systems.

Though we can obtain various higher dimensional local integrable evolution systems in the form \eqref{D+1}, it will be very difficult to find analytical explicit solutions of these integrable models because almost all the traditional methods such as the Hirota's bilinear direct method, Darboux transformations and inverse scattering transformation, etc. can not be directly applied. This fact provides us a challenge even for the fundamental problem, what is integrability? The models obtained by the deformation algorithm of this paper are Lax integrable and symmetry integrable, however, the traditional methods on integrable systems are not valid to find exact solutions even for non-degenerated $n$-soliton solutions for $n\geq 2$.

\begin{figure}
\begin{center}
\subfigure{
\includegraphics[width=0.3\textwidth]{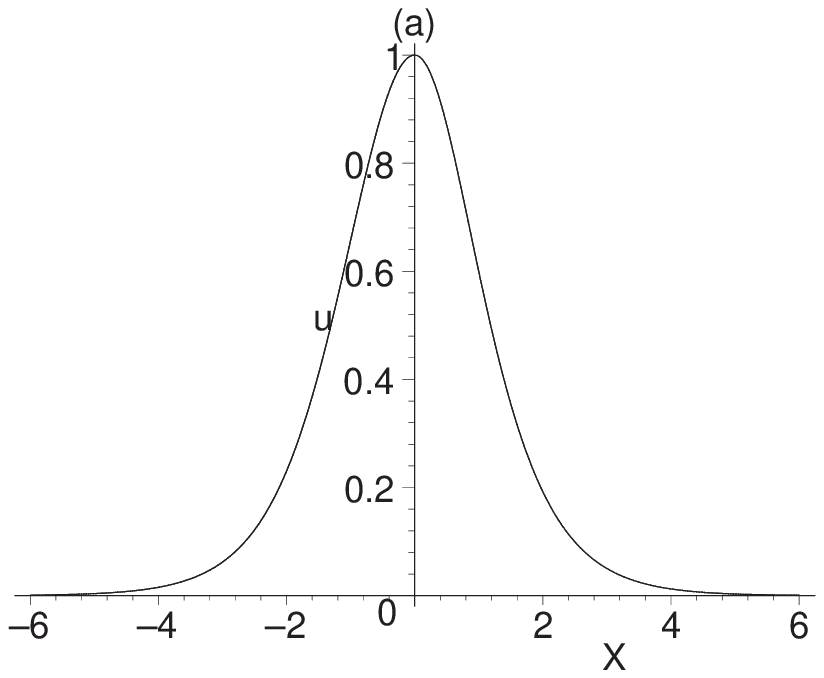}}
\centering
\subfigure{
\includegraphics[width=0.3\textwidth]{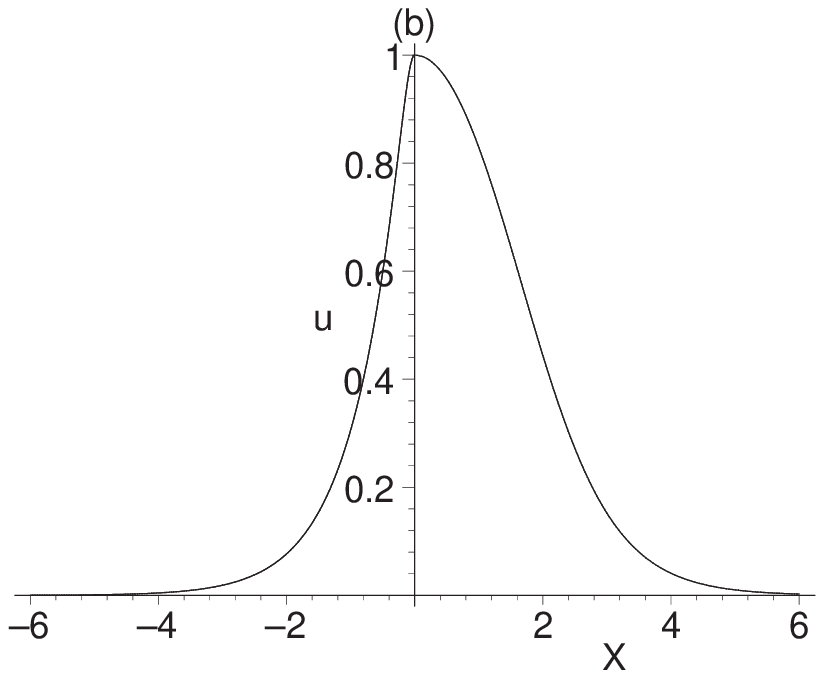}}
\centering
\subfigure{
\includegraphics[width=0.3\textwidth]{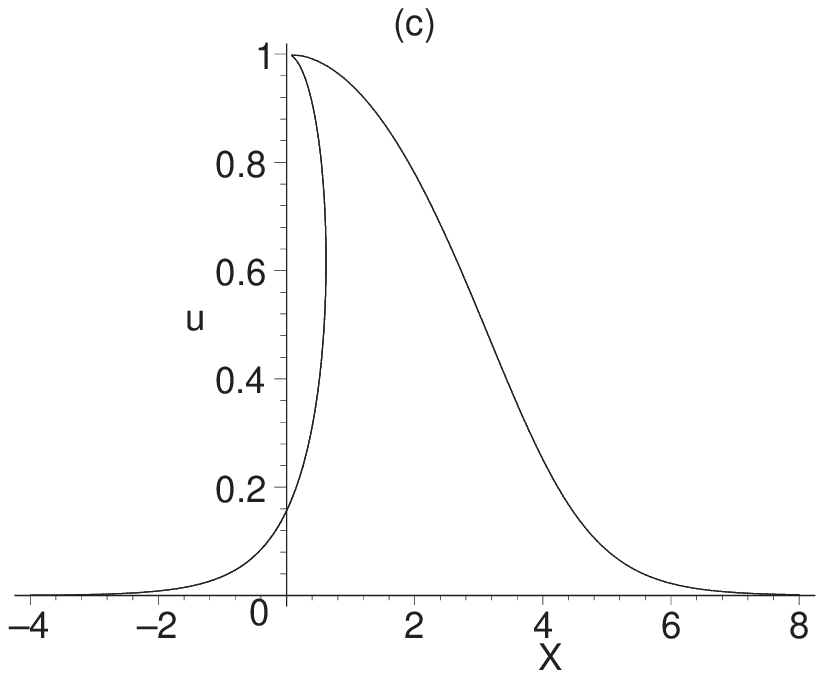}}
\end{center}
\caption{\footnotesize Different single soliton shapes of \eqref{Soliton} with the same wave parameters $\omega=2$ and $k=1$ and different deformation parameters $p$ and $q$. (a). A quasi-symmetric soliton with small deformation parameter selections $p=q=1/30$. (b) An asymmetric soliton with parameter selections $p=q=-1/3$. (c). A foldon under the larger  parameter selections $p=q=-1$ .}\label{case}
\end{figure}

For the travelling wave solution, $u=U(X),\ X=kx+py+qz+\omega t+X_0$, of the (3+1)-dimensional KdV equation \eqref{3+1KdV} satisfies the following ordinary differential equation
\begin{equation}
\omega U_X=\left[(qU^2+pU+k)^3U_{XX}+\frac12U^2(qU^2+2pU+6k)\right]_X+3q(qU^2+pU+k)^2U_X^3.\label{TW}
\end{equation}
The general solution of \eqref{TW} can be expressed by an elliptic integration. The single soliton solution can be implicitly expressed as
\begin{equation}
 U=\frac{\omega}{2k}\mbox{\rm sech}^2\left[\frac{\omega}{6k^3}\sqrt{1-\frac{2k}{\omega}U}
(qkU+3kp+q\omega)+\frac{\sqrt{k\omega}}{2k^2}X\right].\label{Soliton}
\end{equation}
From the soliton expression \eqref{Soliton}, we know that when $p=q=0$, i.e., the soliton is $\{y,\ z\}$-independent, \eqref{Soliton}, it returns back to the standard soliton solution of the usual KdV equation. As $|p|$ and $|q|$ increasing, the soliton shape will deform to an asymmetric soliton  and a foldon or folded solitary wave (multi-valued soliton or solitary wave \cite{foldon}). Fig. \ref{case}a displays the soliton shape expressed by \eqref{Soliton} with small modification parameters $p=q=1/30$ while other parameters are fixed as $\omega=2$ and $k=1$. In this case, the soliton shape is still quite symmetric. Fig. \ref{case}b shows us that the soliton has deformed to an asymmetric one where the parameters are selected as $p=q=-1/3,\ \omega=2$ and $k=1$. For larger selections of the deformed parameters, $p=q=-1$, the soliton becomes a foldon as shown in Fig. \ref{case}c where the values of $\omega$ and $k$ are same as in other two figures.

In the deformation conjecture, we have put some restrict conditions on it. The original model should be a (1+1)-dimensional, local and evolution system and the conserved density should be field derivative independent. From the discussions related to the VAKNS system \eqref{VAKNS}, we know that the conserved density may be field derivative dependent. In fact, other restrict conditions may also be loosed however, this is not the topic of this paper and would be discussed elsewhere.

\section*{Acknowledgement}
The work was sponsored by the National Natural Science Foundation of China (Nos. 12235007, 11975131 and 12275144), K. C. Wong Magna Fund in Ningbo University, Natural Science Foundation of Zhejiang Province No. LQ20A010009. I would also like to thank Profs. R. X. Yao, X. B. Hu, Q. P. Liu, B. F. Feng and D. J. Zhang for their valuable discussions.

\end{document}